# Influence of entropy changes on reactor period

## V. V. Ryazanov

Institute for Nuclear Research, pr. Nauki, 47 Kiev, Ukraine, e-mail: vryazan19@gmail.com

**Highlights**

- The reactor period is considered as a random variable.

- The effect of entropy changes on the reactor period is taken into account.

- The dispersion and other moments of the reactor period are determined.

The period of a nuclear reactor is usually represented as a deterministic quantity. In this article, the reactor period is described as a random variable. An approach based on the first passage time is used. In this case, the effect of the change in entropy on the value of the period of the reactor is determined. Relations are obtained that describe the effect of entropy on the period of the reactor. Expressions are obtained for the dispersion of the reactor period. This value cannot be found from a deterministic approach. Knowing the effect of total entropy change on the period of the reactor makes it possible to control the behavior of the reactor.

## 1. Introduction

The neutron flux in the reactor at different values of the neutron multiplication factor changes at different rates. The parameter of the reactor period - the time during which the neutron flux density changes by a factor of *e* - plays a very important role in characterizing the behavior of the reactor. Its safety depends on the period of the reactor.

Nuclear power plants operate automated control systems that continuously measure the period of the reactor. In works on the kinetics of the reactor (for example, [1-12]), the contributions of the value of neutrons to the dynamics of neutrons [3, 4, 5], the role of the spatial inhomogeneity of the distribution of neutrons [1-12], as well as other effects on which the behavior of the reactor depends are shown.

In this paper, we consider the contribution to the reactor period of effects that have not been studied before, namely, entropy changes during the reactor period. The reactor period is studied using the stochastic approach of the first passage time (*FPT*) of a certain level by a random process [13-15]. Any real physical process is accompanied by a change in entropy. It was shown in [15, 16] that this change in entropy affects the random processes that model the behavior of the physical process, including the first passage time and the moments of this random variable. In this paper, we use the Shannon entropy of the form

$$s_\gamma = -\int P(\Gamma) \ln P(\Gamma) d\Gamma, \qquad (1)$$

where $P(\Gamma)$ is the distribution of $\Gamma$. There are other representations of the entropy functional that satisfy the axioms that entropy must satisfy. This, for example, is the entropy of Kolmogorov-Sinai, Kullback, Tsallis, Renyi, for example, [17]. In [18-20], it was proposed to control the safety of the reactor along with the period of the reactor to use its stochastic analogue - the time to reach a given level by the number of neutrons. The number of neutrons in the reactor is considered as a random process.

In the present work, a simplified point model of a nuclear reactor serves as the starting point. Generalizations of this model are possible. Changes in total entropy include arbitrary effects on the system. Using the proposed approach, it is possible to evaluate the effectiveness of various methods for controlling the behavior of the reactor (movement of rods, etc.), taking into account the exchange of entropy with the environment.

In the theory of nuclear reactors, the period of the reactor is determined from the deterministic equations of neutron kinetics. However, many processes in a reactor (collisions of neutrons with nuclei, fission of nuclei, emission of neutrons, etc.) are of a probabilistic nature. Therefore, it is necessary to involve statistical methods in the study of the reactor period [18-20]. In this work, as in [18-19], the reactor period is considered as a random value of the *FPT* of reaching the level (in this case, it differs by *e* times from the initial level). On the basis of [15], the time of the first reaching is associated with the changes in entropy during this time. In [16], using the example of mesoscopic electron transfer, it was shown that the effect of entropy changes on the process of the *FPT* of reaching a level during this reaching can be significant. Therefore, such a study for the period of the reactor seems to be necessary. The distribution of the first passage time specifically for the reactor period is achieved by introducing a deterministic reactor period into the distribution as a value to the average random *FPT*.

Probabilistic methods have long been used in the theory of nuclear reactors. The neutron flux or their energy is described by the Maxwell distribution [1]. Some theoretical treatment has been given to reactor fluctuations in the literature [21-25]. The theory of branching random processes is used to describe the behavior of neutrons in a reactor [26-29]. Therefore, the use of the *FPT* a certain level by a random process seems natural and logical. This use follows from the fact that the neutron flux is a random process, and *FPT* is one of the properties of this random process. The energy or neutron flux changes with time and in the process of evolution reaches some given level (in our case, a level that differs by *e* times from the initial value).

The period of the reactor is constantly measured. In the theory of nuclear reactors, the period of the reactor is described by deterministic equations for the flux of prompt and concentrations of precursor nuclei of delayed neutrons. In the probabilistic description, these concentration values are their average values. Higher moments of the random variable are also recorded. In a stationary state, changes in a random variable occur due to fluctuations. The stationary state of the reactor coincides with the critical state. In the critical state, the fluctuations are especially large.

In this article, the approach of articles [15-16, 18-20, 30-32] is used, in which *FPT* by a random process is associated with changes in the entropy of the system during this time. The approach is based on the introduction of a statistical distribution containing a random thermodynamic parameter of the *FPT*. The entropy is entered as the averaged logarithm of this distribution (2). The Shannon entropy (1) goes for the changes due to the internal processes in the system and is only a part of the overall entropy and changes of entropy. To define all the changes in the system entropy the exchange with the surroundings should also be accounted. There are opportunities to consider various methods of controlling the behavior of the reactor. The article considers the simplest case of a stationary state without including control of the reactor behavior in the consideration.

In the second section, the reactor period is associated with entropy changes over this period using the results [15] for entropy of the form (1). In the third section, the results obtained are illustrated by calculations. Brief conclusions and discussion are presented in Section 4.

## 2. Statistical description of the reactor period

The lifetime of neutrons in a reactor is finite. In [20], this essential circumstance is used to determine the energy spectrum of neutrons in a reactor. In [20] the neutron spectrum was obtained, which corresponded to the experimentally observed data. In this case, a statistical distribution is used, which contains *FPT* [30, 31] or lifetime in terms of [32].

In [15, 16, 30-32] the statistical distribution of the random variable the *FPT* (or lifetime) $T_\gamma$ as thermodynamical value of the system was obtained; $\gamma$ is thermodynamical parameter conjugate to FPT $T_\gamma$ (2), how the inverse temperature $\theta^{-1} = 1/T_1$ is conjugate to the random energy $u$. The density of such a distribution in the phase space of variables $z$, which are the coordinates and momenta of all particles of the system, has the form

$$\rho(z;u,T_\gamma) = \exp\{-\theta^{-1}u - \gamma T_\gamma\}/Z(\theta^{-1},\gamma), \qquad \theta^{-1} = 1/T_1, \qquad (2)$$

where $u$ is the energy of the neutron system (we assume this value to be a random process); one can use the neutron flux $\Phi$ instead of the energy of the neutron system $u$; $T_1$ is the temperature (in energy units),

$$Z(\theta^{-1},\gamma) = \int \exp\{-\theta^{-1}u - \gamma T_\gamma\}dz = \iint du\, dT_\gamma\, \omega(u,T_\gamma)\exp\{-\theta^{-1}u - \gamma T_\gamma\}, \qquad \theta^{-1} = 1/T_1 \qquad (3)$$

is the partition function, $u$ and $T_\gamma$ are functions of variables $z$ ($z=(q_1,...,q_N; p_1,...,p_N)$), coordinates $q_i$ and momenta $p_i$ of all particles of the system for a system of $N$ particles. The value $\theta^{-1}$ is equal to the reciprocal temperature $1/T_1$, and $\gamma$ as $\theta^{-1}$ are determined from the equations for the averages:

$$\langle u \rangle = -\partial \ln Z / \partial \theta^{-1}\big|_\gamma; \qquad \langle T_\gamma \rangle = -\partial \ln Z / \partial \gamma\big|_{\theta^{-1}}, \qquad \theta^{-1} = 1/T_1, \qquad (4)$$

where brackets denote averaging. Here we have introduced cells of the "extended" phase space [15, 18-20, 30-32] with constant values ($u$, $T_\gamma$) (we have replaced the phase cells with constant values of $u$). Instead of the value of the structure factor $\omega(u)$, the value $\omega(u,T_\gamma)$ is introduced, which describes the volume of the hypersurface in the phase space, with fixed values of the energy $u$ and FTP $T_\gamma$. The parameter $\mu(u, T_\gamma)$ is equal to the number of phase space states with parameters less than $u$ and $T_\gamma$, and $\omega(u, T_\gamma) = d^2\mu(u, T_\gamma)/du\, dT_\gamma$. In this case $\int \omega(u, T_\gamma)dT_\gamma = \omega(u)$. Function $\omega(u, T_\gamma)$ corresponds to a random process, and this function is the joint probability density of an unperturbed process without external influences in a stationary state. The factor $\omega(u,T_\gamma)$ is the joint probability for $u$ and $T_\gamma$, considered as the stationary probability of this process. In this case, the distribution (2)-(3) generalizes the Gibbs distribution, in which $\gamma=0$, describes stationary non-equilibrium systems, including the ensemble of neutrons in the reactor.

The joint distribution density of the macroscopic parameters, random variables $u$ and $T_\gamma$ is equal to

$$p(u,T_\gamma) = \int \delta(u - u(z))\delta(T_\gamma - T_\gamma(z))\rho(z;u(z),T_\gamma(z))dz =$$
$$= \rho(u,T_\gamma)\omega(u,T_\gamma) = \exp\{-\theta^{-1}u - \gamma T_\gamma\}\omega(u,T_\gamma)/Z(\theta^{-1},\gamma). \qquad (5)$$

Integrating (5) over FPT $T_\gamma$, we obtain

$$p(u) = \int p(u,T_\gamma)dT_\gamma = \frac{e^{-u/\theta}}{Z(\theta^{-1},\gamma)}\int_0^\infty \omega(u,T_\gamma)e^{-\gamma T_\gamma}dT_\gamma, \qquad \theta^{-1} = 1/T_1. \qquad (6)$$

The form of the function $\omega(u,T_\gamma)$ must correspond to the assumptions written above. Therefore, we write the functions $\omega(u,T_\gamma)$ in the form

$$\omega(u,T_\gamma) = \omega(u)\omega_1(u,T_\gamma) = \omega(u)\sum_{i=0}^n P_i f_i(T_\gamma,u). \qquad (7)$$

Expression (7) is written in general form. It is assumed that the system has $n + 1$ classes of states, $P_i$ is the probability that the system is in the $i$-th class of states, $f_i(T_\gamma,u)$ is the probability density that for a system from the $i$-th class the first passage time is equal to $T_\gamma$. Setting the function (7) determines the correspondence of the distribution parameter $T_\gamma$ in (2) to the period of the nuclear reactor.

Consider the system of neutrons in a nuclear reactor as a statistical system. Prompt neutrons are considered as a zero class, and $n$ groups of delayed neutrons are considered as $n$ classes. If we choose the function $f_i$, for example, in (7) in the form of a gamma distribution

$$f_i(x) = \frac{1}{\Gamma(\alpha_i)} \frac{1}{b_i^{\alpha_i}} x^{\alpha_i-1} e^{-x/b_i}, \quad x>0, \quad f_i(x)=0; \quad x<0; \quad \int_0^\infty e^{-\gamma_i x} f_i(x) dx = (1+\gamma_i b_i)^{-\alpha_i}, \quad (8)$$

where $\Gamma(\alpha)$ is the gamma function, $b_i$, and put $b_i \alpha_i = T_{\gamma 0i}$, ($T_{\gamma 0i} = T_{i\gamma=0}$ are the unperturbed average lifetimes of the system in the $i$-th states [15, 16, 30-32]), $\alpha_i = \gamma_i/\lambda_i$, where the parameter $\lambda_i$ is equal to the intensity of entry into the $i$-th subsystem, $\gamma_i = \gamma$ in the $i$-th subsystem, then from (5)-(8) we obtain that

$$(1+\gamma_i b_i)^{-\alpha_i} = (1+\lambda_i T_{\gamma 0i})^{-\gamma_i/\lambda_i}; \quad p(u) = \int p(u,T_\gamma) dT_\gamma = \frac{e^{-\theta^{-1}u} \omega(u)}{Z(\theta^{-1},\gamma)} \sum_{i=0}^n P_i/(1+\lambda_i T_{\gamma 0i})^{\alpha_i}. \quad (9)$$

If $\alpha_i=1$, $\gamma_i=\lambda_i$, $b_i=T_{\gamma 0i}$, $f_i$ is an exponential distribution. For the exponential distribution in (8)-(9) and one class of ergodic states when $n=0$,

$$(1+\gamma T_{\gamma 0})^{-1} = <T_\gamma>/T_{\gamma 0}, \quad (10)$$

where $<T_\gamma>$ is the average lifetime of the system obtained from (3)-(4), (7)-(8) [15, 16, 30-32], we have a situation where only prompt neutrons are taken into account. For neutrons $<T_\gamma> = |1/\omega|$, where $|1/\omega| = T$ is the reactor period [1, 2];

$$<T_\gamma> = l/|k-1|, \quad \gamma l = |k-1|-1, \quad T_{\gamma 0} = l. \quad (11)$$

In (11), the quantity $l = l_{ef}/(1+L^2 B^2)$ describes the lifetime of neutrons in the reactor. In this case, the value $l_{ef} = 1/v\Sigma_a \sim 10^{-3}$ sec is equal to the average effective neutron lifetime in a nuclear reactor ($v$ is the average neutron velocity, $\Sigma_a$ is the neutron absorption cross section in the reactor), $L^2$ is the square of the diffusion length, $B^2$ is the geometric Laplacian [1, 2], $k$ is the effective neutron multiplication factor [1–12].

The quantities

$$<T_{\gamma i}> = T_{\gamma 0i}(1+\gamma_i T_{\gamma 0i})^{-1} = |1/\omega_i| = T_i \quad (12)$$

are sought as solutions to the inverse clock equation [1, 2], $T_i$ is the reactor period. If we consider one effective group of delayed neutrons, then the inverse clock equation has the form [1-4]

$$k\rho(\omega+\lambda) = \omega l_{ef}(\omega+\lambda) + \beta\omega k, \quad (13)$$

where $\rho=(k-1)/k$ is the reactor reactivity. We assume $\lambda=0.077$ sec$^{-1}$ is the decay constant of delayed neutron precursors, $\beta=0.0065$ is the fractional yield of the delayed neutrons. The solution of equation (13) has the form

$$\omega_{0,1} = (p/2l)[\pm C - 1], \quad l = l_{ef}, \quad p = \lambda l + 1 - k + \beta k, \quad C = (1+4\lambda l(k-1)/p^2)^{1/2}. \quad (14)$$

We assume in (9) $P_0=(1-\beta)$, $P_1=\beta$ are the probabilities that neutrons are in the classes of prompt or delayed neutrons, $T_{00\ lifetime}=l$, $T_{01\ lifetime}=1/\lambda$ are unperturbed average lifetimes of prompt and delayed neutrons, from (6), (9), (14) we obtain that in the case of one group of delayed neutrons

$$\rho(u) = \exp\{-\theta^{-1}u\} Z^{-1} \omega(u) [2(1-\beta)/(\lambda l + \beta k + 1 - k)(1-C) + \\ + 2\beta\lambda l/(\lambda l + \beta k + 1 - k)(1+C)]; \quad C = [1+4\lambda l(k-1)/(\lambda l + \beta k + 1 - k)^2]^{1/2}. \quad (15)$$

Several groups of delayed neutrons are considered in the same way. In (7), (15) we take $\omega(u) \sim u^{1/2}$. Specifying the distribution density $f_i(T_\gamma)$ in (7) in the form of an exponential distribution for the *FPT* of the period of the reactor of the form (12), (13), obtained from the deterministic approach in [1, 2], determines the random value of the *FPT* by the neutron flux of a level $e$ times different from the initial period of the reactor.

We will look for a partition function of the form (3). We will use expressions (3)-(4), and also assume that the distribution of the first passage time is independent of the random energy value (dependence on the average energy is possible). Then the variables are separated, the integration is carried out over independent variables, and as in [15]

$$Z(\theta^{-1},\gamma) = Z_{\theta^{-1}} Z_\gamma, \quad Z_{\theta^{-1}} = \int e^{-\theta^{-1}u} \omega(u) du, \quad Z_\gamma = \int_0^\infty e^{-\gamma T_\gamma} \sum_{j=0}^n P_j f_j(T_\gamma) dT_\gamma, \quad \theta^{-1} = 1/T_1. \quad (16)$$

Let us normalize the distribution for $T_\gamma$, making it dimensionless. In order for $P(T_\gamma)$ and $Z_\gamma$ to be dimensionless, we introduce the times $\tau_0$ and $\tau_1$. The distribution for *FPT* $T_\gamma$ takes the form

$$P(T_\gamma) = \frac{e^{-\gamma T_\gamma}}{Z_\gamma}[(1-\beta)(-\tau_0 \omega_0 / \bar{\tau})e^{\omega_0 T_\gamma} + \beta(-\tau_1 \omega_1 / \bar{\tau})e^{\omega_1 T_\gamma}]. \qquad (17)$$

With such a record, the probabilities that the system is in the *i*-th class of states (prompt or delayed neutrons), from (7), (9), (16) are not $P_0=(1-\beta)$, $P_1=\beta$, and

$$P_0 = (1-\beta)\tau_0/\bar{\tau}, \quad P_1 = \beta\tau_1/\bar{\tau}, \quad \tau_0 = l, \quad \tau_1 = 1/\lambda, \quad \bar{\tau} = (1-\beta)\tau_0 + \beta\tau_1. \qquad (18)$$

This corresponds to the times $T_{00lifetime}=l$, $T_{01lifetime}=1/\lambda$ that neutrons stay in the states after their instantaneous appearance during nuclear fission or with a delay after their emission by precursor nuclei, taking into account the fractions of those and other neutrons.

From the normalization condition of the distribution (17) we have

$$\int P(T_\gamma)dT_\gamma = \frac{1}{Z_\gamma}[(1-\beta)(-\tau_0\omega_0)/(\gamma-\omega_0)\bar{\tau} + \beta(-\tau_1\omega_1)/(\gamma-\omega_1)\bar{\tau}] = 1, \quad \gamma-\omega_0 > 0, \quad \gamma-\omega_1 > 0,$$

and we determine the partition function of the *FPT* in the form

$$Z_\gamma = [(1-\beta)(-\tau_0\omega_0)/(\gamma-\omega_0)\bar{\tau} + \beta(-\tau_1\omega_1)/(\gamma-\omega_1)\bar{\tau}], \quad \bar{\tau} = (1-\beta)\tau_0 + \beta\tau_1. \qquad (19)$$

Comparing relations (5), (7), (16), (17), (18), (19), we obtain that

$$\sum_{j=0}^{n} P_j f_j(T_\gamma) = [(1-\beta)(-\tau_0\omega_0/\bar{\tau})e^{\omega_0 T_\gamma} + \beta(-\tau_1\omega_1/\bar{\tau})e^{\omega_1 T_\gamma}] = \omega(T_\gamma), \qquad (20)$$

$$\omega(u,T_\gamma) = \omega(u)\omega_l(u,T_\gamma) = \omega(u)\sum_{i=0}^{n} P_i f_i(T_\gamma) = \omega(u)\omega(T_\gamma); \quad \sum_{i=0}^{n} P_i f_i(T_\gamma) = \omega(T_\gamma).$$

The change in entropy over the period of the reactor or during the *FPT* is expressed in terms of the distribution parameter $\gamma$ from (2) using relation (1). Entropy of the distribution (2) $s = -\langle \ln \rho(z,u,T_\gamma)\rangle$ describes the internal entropy of the system (brackets mean averaging). This entropy, in accordance with (16), consists of two parts:

$$s = -k_B \int p(u,T_\gamma)\ln[p(u,T_\gamma)]dudT_\gamma = s_{\theta^{-1}} + s_\gamma, \quad s_{\theta^{-1}} = \theta^{-1}<u> + \ln Z_{\theta^{-1}}, \quad s_\gamma = \gamma<T_\gamma> + \ln Z_\gamma. \qquad (21)$$

These expressions are obtained from relations (2), (16), (17). The same expressions are also written from distribution (5). Variables are separated as in (16). From (7) we obtain $\omega(u,T_\gamma) = \omega(u)\sum_{j=0}^{n} P_j f_j(T_\gamma)$, $\omega(T_\gamma) = \sum_{j=0}^{n} P_j f_j(T_\gamma)$, as in (20).

The value of (5) is $p(u,T_\gamma) = (e^{-\theta^{-1}u}\omega(u)/Z_{\theta^{-1}})(e^{-\gamma T_\gamma}\sum_{j=0}^{n} P_j f_j(T_\gamma)/Z_\gamma) = p(u)p(T_\gamma)$, where $Z_{\theta^{-1}}$, $Z_\gamma$, from (16). From expressions (1) and (5) we obtain expression (21). In this case $s = s_\gamma + s_{\theta^{-1}}$ (*s* is the entropy density), and

$$s = s_\gamma + s_{\theta^{-1}} = s_{\theta^{-1}} - \Delta = \gamma \bar{T}_\gamma + \theta^{-1}\bar{u} + \ln Z = \theta^{-1}\bar{u} + \ln Z_{\theta^{-1}} - \Delta, \quad -\Delta = s_\gamma = s - s_{\theta^{-1}}. \qquad (22)$$

Expression (22) follows from relations (1)-(2). The part of the entropy that depends on the energy is $s_{st} = s_{\theta^{-1}} = \theta^{-1}\bar{u} + \ln Z_{\theta^{-1}}$. The same expressions are obtained in Appendix A using the approaches of [33]. Let us rewrite (22) in the form

$$-\Delta = s_\gamma = \gamma \bar{T}_\gamma + \ln Z_\gamma. \qquad (23)$$

The quantities included in (23) are written in (2)-(4), (10)-(12), (19), $\Delta \geq 0$.

Note that along with the Shannon entropy (1), the Renyi and Tsallis entropy [17] and fractal entropy can be used.

Expressions (20)-(22) includes the average the first passage time (4), (19), which we consider equal to the period of the reactor, taking into account the change in entropy $\Delta$ (22), which is related to the conjugate to *FPT* parameter $\gamma$ from distribution (2) by relation (23). From (4), (19) for the average period of the reactor we have

$$\bar{T}_\gamma = \int T_\gamma P(T_\gamma)\frac{dT_\gamma}{\bar{\tau}} = \frac{(1-\beta)T_0(\tau_0/\bar{\tau})(1+\gamma T_1)^2 + \beta(\tau_1/\bar{\tau})T_1(1+\gamma T_0)^2}{[(1-\beta)(\tau_0/\bar{\tau})(1+\gamma T_1) + \beta(\tau_1/\bar{\tau})(1+\gamma T_0)](1+\gamma T_0)(1+\gamma T_1)}, \quad (24)$$

where according to (12), (14), (18)

$$T_0 = \left|\frac{1}{\omega_0}\right| = \left|\frac{2l}{p(1-C)}\right|, \quad T_1 = \left|\frac{1}{\omega_1}\right| = \left|\frac{2l}{p(1+C)}\right|, \quad x = \tau_1/\tau_0, \quad x = \frac{1}{\lambda l},$$

$T_0$ and $T_1$ are the steady and transition periods of the reactor [1, 2].

The period of the reactor (24) at $\gamma=0$ is equal to

$$\bar{T}_{\gamma=0} = T_{00} = [(1-\beta)T_0 + \beta x T_1]/(1-\beta+\beta x). \quad (25)$$

It is necessary to determine the parameter $\Delta$, the internal entropy of the reactor. In [34-36], various approaches to determining the entropy of a reactor of the form (1) are considered. Thus, in [34], a methodology based on sample entropy is presented. This methodology was used to develop reactor stability monitoring. This methodology was tested on a set of signals from several operating nuclear power plants that experienced various kinds of unstable events. The resulting values of the Shannon entropy of this *BWR* (boiling water reactor) signal give average values of the order of 0.47. The approach considered in [36] is aimed at revealing the ways of isotope accumulation during the entire fuel evolution. The obtained value of the Shannon entropy for U238 is 0.19. If we take the average value of these two entropies, we get that $\Delta \approx 0.33$.

Similar results are obtained by using an estimator based on the approach of extended non-equilibrium thermodynamics [37], when entropy production is proportional to the square of the neutron flux. Note that the calculations show a weak dependence of the results on the value of the parameter $\Delta$. For example, at $\Delta=0.3$ and $\Delta=0.44$, $\Delta=0.7$, the results differ slightly.

Let us find the parameter $\gamma$ from relation (23). Expanding the right side of (23) in powers of $\gamma$ up to the second order, we obtain the relation

$$-\Delta \approx \frac{1}{2}\gamma^2 \frac{\partial \bar{T}_\gamma}{\partial \gamma}\bigg|_{\gamma=0}, \quad \gamma \approx [-2\Delta / \frac{\partial \bar{T}_\gamma}{\partial \gamma}\bigg|_{\gamma=0}]^{1/2}. \quad (26)$$

where $\partial \bar{T}_\gamma / \partial \gamma|_{\gamma=0} = -D_{T_\gamma|\gamma=0}$ - *FPT* variance at $\gamma=0$.

The expression for the reactor period, taking into account only prompt neutrons, can be obtained by tending beta to zero in the resulting expression (24);

$$\bar{T}_\gamma = \frac{(1-\beta)T_0(1-\lambda l(1-k))^2 + \beta \lambda l(1-k)T_1(1-p^2)^2/2p^2}{[(1-\beta)(1-\lambda l(1-k)) + \beta \lambda l(1-k)(1-p^2)/2p^2](1-p^2)(1-\lambda l(1-k))}\bigg|_{\beta=0} \approx \frac{T_0}{1-p^2}, \quad (27)$$

$$\bar{T}_\gamma(\beta=0) = \left|\frac{T_0}{1-p_0^2}\right|, \quad p_0 = p(\beta=0) = 1-k+\lambda l.$$

The bar above the symbol indicates averaging.

Perhaps these results will change when using a non-exponential distribution, for example, a) gamma distribution (8), when an additional parameter is found from the distribution variance; b) expansion into a series, taking into account an additional term, etc.

The proposed approach makes it possible to find fundamentally new moments in the description of the behavior of the reactor. Thus, the period of the reactor is usually considered in a deterministic way. Representation of the reactor period as a random variable of the *FPT* makes it possible to determine not only the average value of the reactor period, but also arbitrary moments of this quantity, which is already considered as a random variable. So, from the expression (19) we find the second moment of the reactor period and the dispersion of the reactor period in the form

$$D_{T_\gamma} = \langle T^2_\gamma \rangle - \langle T_\gamma \rangle^2 = \partial^2 \ln Z_\gamma / \partial \gamma^2 = 2[(1-\beta)(-\tau_0\omega_0)/(\gamma-\omega_0)^3 \bar{\tau} + \beta(-\tau_1\omega_1)/(\gamma-\omega_1)^3 \bar{\tau}]/Z_\gamma - $$
$$-([(1-\beta)(-\tau_0\omega_0)/(\gamma-\omega_0)^2 \bar{\tau} + \beta(-\tau_1\omega_1)/(\gamma-\omega_1)^2 \bar{\tau}]/Z_\gamma)^2 \quad (28)$$

Substituting into this relation the obtained expression for the parameter $\gamma$, which is written in approximate form in (26), leads to taking into account those changes in entropy in the system that occur during the period of the reactor. It is possible to write more precise expressions for $\gamma$.

## 3. Results

Calculations using the MatLab computational code using expressions (12)-(26) give the following results of $<Tg(k)> = \bar{T}_\gamma$ for various values of $k$, Fig. 1-3. The red color shows the dependences on $k$ of the reactor period, taking into account changes in entropy $<Tg(k)> = \bar{T}_\gamma$ (24); in blue, the dependences of the reactor period on $k$ without taking into account entropy changes, in the form $\bar{T}_{\gamma=0} = T_{00} = [(1-\beta)T_0 + \beta T_1]/(1-\beta+\beta x)$ (25).

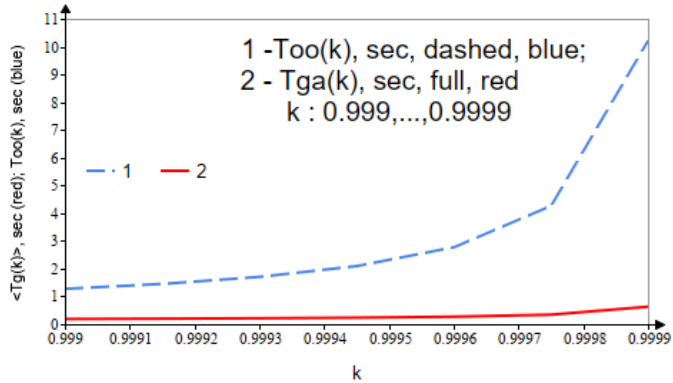

Fig. 1. Behavior of the reactor periods $<Tg(k)> = \bar{T}_\gamma$ (24) (in red) taking into account changes in entropy over this period and the reactor period $T_{00}(k) = \bar{T}_{\gamma=0}(k)$ (25) (in blue) without such consideration in the interval $k$: 0.999,…, 0.9999.

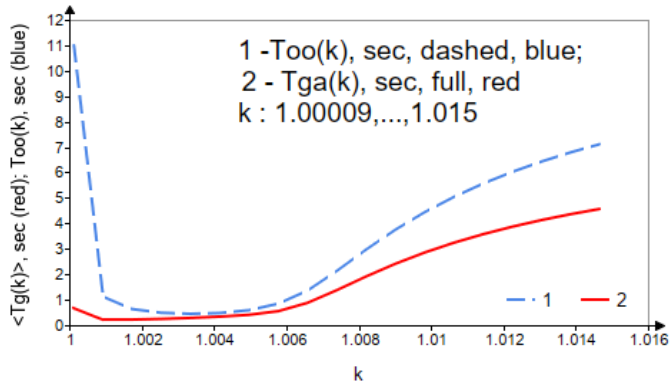

Fig. 2. The same as in Fig. 1, in the interval $k$: 1.00009, …, 1.015.

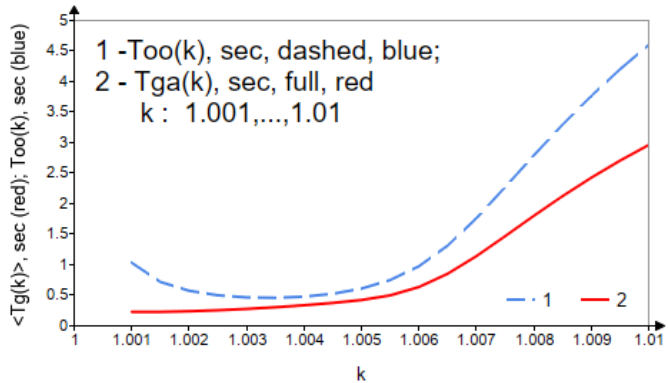

Fig. 3. The same as in Fig. 1, in the interval $k$: 1.001, …, 1.01.

Let us compare the obtained results with the known approximations for the cases $k>1$, when $\rho = (k-1)/k \ll \beta$ and $\rho = (k-1)/k \gg \beta$, $\beta=0,0065$, obtained in [1, 2]. To do this, we substitute the approximate expressions for $T_0$ and $T_1$ written in [1, 2] into relation (24), obtaining the relation for $<Tg1(k)>$. On Fig. 4 shows such a comparison for $\rho \gg \beta$ when the parameter $k$ changes in the interval 1.07,…, 1.09. The values $T_0$ and $T_1$ obtained in [1, 2] for this case are equal to $T_0 = |lk/(k-1)|$, $T_1 = |-1/\lambda|$, $l \approx 10^{-3}$ sec, $\lambda \approx 0.077\,\mathrm{sec}^{-1}$. Substituting these values into expression (24), we obtain the dependence $<Tg1(k)>$, shown in green in Fig. 4. This relationship is compared with (24) (in red) taking into account changes in entropy over this period (the expressions for $T_0$ and $T_1$ are taken from the relationships written below (24)) and the reactor period (25) (in blue) without such consideration in the interval $k$: 1.07, …, 1.09. It can be seen that the $<Tg1(k)>$ value is closer to the dependence $<Tg(k)> = \overline{T}_\gamma$ than to $T_{00}(k) = \overline{T}_{\lambda|\gamma=0}(k)$.

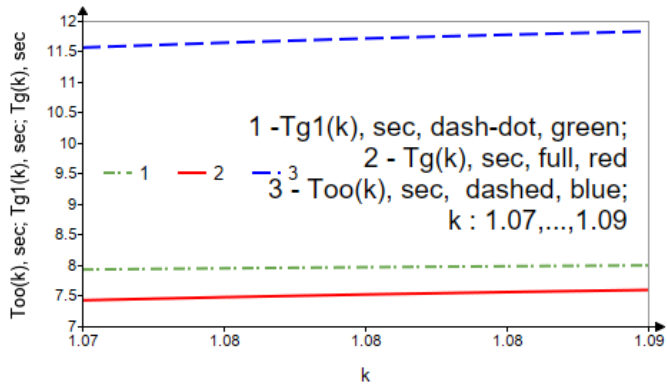

Fig. 4. For the case $\rho = (k-1)/k \gg \beta$, in the interval $k=1.07$, …, 1.09, the results calculated for $T_{00}(k) = \overline{T}_{\lambda|\gamma=0}(k)$, blue, $<Tg(k)> = \overline{T}_\gamma$, red, are compared with the approximations obtained for this case in [1], where the values $T_0 = |lk/(k-1)|$, $T_1 = |-1/\lambda|$, $l \approx 10^{-3}$ sec, $\lambda \approx 0.077\,\mathrm{sec}^{-1}$, are substituted into relation (24) and the expression $<Tg1(k)>$, green, is obtained.

For the case $\rho = (k-1)/k \ll \beta$, consider the behavior of the average period of the reactor in the interval $k=1.0005,…,10013$, assuming that $0.0005$, $0.0013 \ll 0.0065$. The approximations $T_0 = (\beta - \rho + \lambda l)/\lambda \rho$, $\beta = 0.0065$, $T_1 = |-l/(\beta - \rho + \lambda l)|$ written in [2] for this case are substituted into expression (24). Calculation using the resulting expression leads to complete agreement with $<Tg(k)> = \overline{T}_\gamma$.

On the Figures 5 - 6 show the behavior of the dispersion of the reactor period (28) in the different intervals of the multiplication factor $k$. The reactor period is considered as a random variable. The dispersion of this random variable is not equal to zero both for $\gamma>0$ and for $\gamma=0$ and depends on $k$. The value $Dg0(k) = D_{T_\gamma|\gamma=0}$ represents the dispersion of the reactor period without taking into account the change in entropy. This quantity appears in relation (26). Behavior $Dg(k) = D_{T_\gamma}$ (28) is shown in red, behavior $Dg0(k) = D_{T_\gamma|\gamma=0}$ is shown in blue.

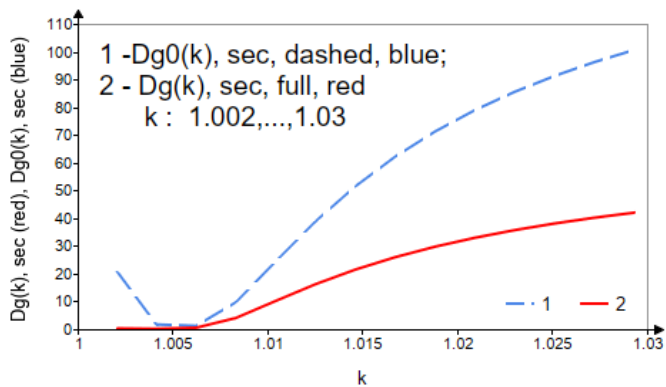

Fig. 5. The behavior of the reactor period dispersion $Dg(k) = D_{T_\gamma}(k)$ (28) (in red) taking into account changes in entropy over this period and the reactor period dispersion $Dg0(k) = D_{T_\gamma|\gamma=0}(k)$ (in blue) without such consideration in the interval $k$: 1.002,…, 1.03.

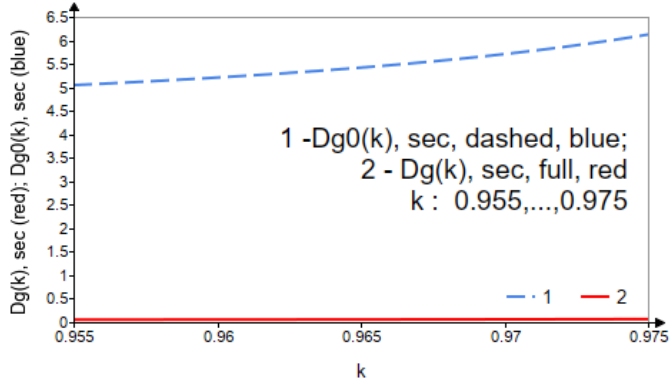

Fig. 6. The same as in Fig. 5, in the interval $k$: 0.955,…, 0.975.

From a comparison of Fig. 3 and 5 it can be seen that, for example, for $k$=1.004, the relationship between the mean value $\bar{T}_g(k) = \bar{T}_\gamma(k)$ and the dispersion $D_{T_\gamma}$ is as follows: $\bar{T}_g(k) = \bar{T}_\gamma(k = 1.004) = 0.246 \gg D_{T_\gamma}(k = 1.004) = 0.061$. This corresponds to the exponential distribution when $\bar{T}_\gamma(k) = 1/\lambda$, $D_{T_\gamma} = 1/\lambda^2$, and $D_{T_\gamma} = 1/\lambda^2 \approx 0.061 \approx (\bar{T}_\gamma(k))^2 = (0.246)^2 = 0.06052$. A similar situation for the gamma distribution (8), when $\bar{T}_\gamma(k) = \alpha/\lambda$, $D_{T_\gamma} = \alpha/\lambda^2$. Possible errors in the measurement of the reactor period depend on many factors. The root mean square error is equal to the square root of the variance. The proposed approach predicts a root-mean-square error in determining the period of the reactor, equal to one hundred percent. For example, in [38] the conditions under which this is true are indicated. So, with ratio of counting time to total measurement time, $t_0/t < 1.5$, accuracy at given reliability > 100 percent. In [38], the possibilities of increasing the accuracy up to 10 percent and higher are indicated.

## 4. Conclusions and discussion

A power reactor with feedback is not described by an asymptotic period in accordance with the inverse clock equation - such a period cannot be steady-state and therefore an unambiguous relationship between the reactor period and the introduced reactivity is not determined. And as a consequence, it is difficult to find out how dangerous a change in the effective neutron multiplication factor is. Therefore, the deterministic period of the reactor, widely used in monitoring and control, can reasonably be used only at the minimum controlled power.

It can be seen from the obtained results that taking into account changes in entropy leads to a decrease in both the period of the reactor and its dispersion. Since in real physical systems in non-equilibrium processes there are always changes in entropy, the above estimates should be taken into account when calculating the period of the reactor. This is important for the safety of the reactor.

The paper considers the simplest case of stationary operation of the reactor. However, it is possible to consider other situations that are important in the processes of reactor operation. If we consider the total change in entropy, adding to the entropy generated inside the reactor the exchange of entropy with the environment, then it becomes possible to evaluate various types of external influences on the operation of the reactor. Then further important prospects for taking into account the change in entropy over the period of the reactor will be associated with the possibilities of controlling the period of the reactor. For example, for the most common method of reactor control - control rods - it is possible to calculate the reactor period at any position of the rods and during their movement. Such calculations will allow you to choose the optimal and safe control strategy. The same applies to the use of boric acid, the movements of reflectors, changes in temperature, pressure and other thermodynamic parameters. In addition, any other possibilities of influencing the reactor parameters (mechanical, acoustic, optical, electromagnetic, neutron, etc.) are also evaluated using this approach. The proposed approach makes it possible to choose the most effective control method. However, for this it is necessary to consider the exchange of entropy with the environment, and not just changes in entropy within the system, as for the Shannon entropy (1). This task requires a separate detailed study. Estimation of the average time for the number of neutrons in a reactor to reach a level and the dispersion of this value can serve as one of the practical applications of the proposed description for various options for changing reactivity over time. These changes are assumed to be given. Features of the behavior of the moments in time when the level is reached will indicate the instability of the neutron system.

To prove the correctness of the paper's conclusions, a very detailed Monte Carlo code model can be used, from which it is possible to extract "experimental"-like data. Any indirect argument in favor of the proposed approach can be the fact that the main relations (22)-(23) were obtained in two ways: in the article itself and in Appendix A. Probably the most convincing evidence in favor of the validity of the results obtained is the obvious fact that entropy increases in all irreversible processes, including the *FPT* process. Both the parameter $\theta^{-1} = 1/T_1 \neq 0$ in distribution (2) and the parameter $\gamma \neq 0$ in (2). This thermodynamic parameter $\gamma$ acts as a parameter of the Laplace transform *FPT*, coinciding with the nonequilibrium partition function $Z_\gamma$ (16), and in this relation and in its derivatives, it is also not equal to zero. This parameter $\gamma$ is associated with the change in entropy $\varDelta$ using relations (22), (23). Its further substitution into relation (24) gives an expression for the average period of the reactor through the change in entropy $\varDelta$. For mesoscopic charge transport this is shown in [16].

In real operating reactors, the difference between the periods of the reactor, with and without taking into account the influence of entropy changes, is apparently not as large. The accuracy of determining the period of the reactor is also apparently greater than the 100 percent root-mean-square error predicted by the stochastic theory in this article. In [38] noted that an accuracy of 10 percent or more is achievable. The reason for this is apparently the effects of reactor control. These effects are deterministic, they operate in both stationary and transient modes, suppressing fluctuations and reducing randomness. Time dependencies become correlated. There are both random and deterministic elements in the reactor period. This might reflect a need to alter the approach with random variables. This article reflects only one, the stochastic part of the problem, and does not touch upon its equally important deterministic part. It is possible to pose and solve a separate problem of assessing the contributions of the stochastic effects discussed in this article and the deterministic part of the control. Another reason for the large difference in the results with and without entropy influence can be a rough assumption (16) about the independence of random values of the energy $u$ and *FPT* $T_\gamma$. Therefore, a computational experiment may not give an unambiguous answer about the accuracy of the results obtained.

These circumstances can affect the decrease in the entropy generated in the reactor over the reactor period, the decrease in the difference between $\bar{T}_\gamma(k)$ and $T_{00}(k) = \bar{T}_{|\gamma=0}(k)$, and the accuracy of the measurement of the reactor period. They do not apply to comparison with the results of [1, 2] (Fig. 4), since in [1, 2], in the approximations for the reactor period, the processes of reactor control were also not taken into account. Comparison, apparently, should be carried out either with the measured values of the period of the reactor or with calculated values of the form $T(t) = \Phi(t)/(d\Phi(t)/dt)$.

The time it takes for the number of neutrons to reach a level has a more general physical meaning than the period of the reactor. The reactor period is a dynamic quantity, the average time to reach the level is stochastic, obtained from the stationary distribution of the time of first reaching the level. The period of the reactor is determined by the dynamics of the system. The time it takes for the number of neutrons to reach a level also characterizes the time interval in the evolution of the neutron system, however, the approach to determining this quantity and its physical meaning differ from the dynamic approach used in studying the reactor period. Time-dependent reactivity can be included in expressions for time to reach level. The very definition of the reactor period of this article, such a definition can be called stochastic, differs from the traditional definition of the reactor period of [1, 2]. The stochastic period of the reactor (24) is defined as average *FPT* for a process with a random neutron flux (or energy). The traditional deterministic definition of reactor periods [1-2] is present in this definition. In relation (24), written for one group of delayed neutrons, steady $T_0$ and transition $T_1$ periods of the reactor act as already known parameters from [1, 2]. The situation is the same for six groups of delayed neutrons. In this case, the change in the neutron flux occurs under the influence of all periods of the reactor ($T_0$ and $T_1$ for one group of delayed neutrons), appearing in the definition of the stochastic period of the reactor with different statistical weights. In this article, the influence of entropy change in the reactor on the period of the reactor is determined. This change in entropy is related by relation (26) to the parameter $\gamma$, which is present in distribution (2) as a thermodynamic force conjugate to a random value of the thermodynamic parameter *FPT*.

Note that the proposed approach makes it possible to determine the dispersion of the reactor period and the effect of entropy changes on it. Thus, the stochastic description of the reactor period opens up new possibilities in the theory of nuclear reactors.

## Appendix A. Entropy of macroscopic parameters (energy and FPT)

In [33], issues of nonequilibrium thermodynamics of open systems, systems in which non-zero flows of arbitrary signs are supplied from the external environment, are considered. In [33] the relation

$$s = s_R + \langle s(R) \rangle, \quad s_R = -k_B \int p(R) \ln[p(R)] dR, \quad \langle s(R) \rangle = -k_B \int p(z) \{\ln[p(z)] - \ln[\omega(R(z))]\} dz \quad (A.1)$$

was obtained, where $\omega(R)$ is the distribution density from expression (20), $R$ is a function depending on the dynamic variables $z$; in our case, the $R$ value is a two-component value: $R_1 = u$, $R_2 = T_\gamma$. Expanding the resulting expression for the entropy in a power series in powers of $\gamma$, for small values of $\gamma$ we obtain $s_\gamma = -\gamma^2(\langle T_0^2 \rangle - \langle T_0 \rangle^2) \leq 0$, $s \to s/k_B$, (the entropy is divided by $k_B$, Boltzmann's constant). The terms $\int \frac{e^{-\theta^{-1}u} \omega(u)}{Z_{\theta^{-1}}} \ln[\omega(u)] du$, $\int \frac{e^{-\gamma T_\gamma} \sum_{j=0}^{n} P_j(T_\gamma) f_j(T_j)}{Z_\gamma} \ln[\sum_{j=0}^{n} P_j(T_\gamma) f_j(T_\gamma)] dT_\gamma$ cancel ($\sum_{i=0}^{n} P_i f_i(T_\gamma) = \omega(T_\gamma)$) and the variables are separated, as in (16), in contrast to [20], where this dependence is taken into account. The relation (22) is written. The Shannon entropy (1) for the distribution of *FPT* is equal $s_{T_\gamma} = \gamma \bar{T}_\gamma + \ln Z_\gamma + s^\beta_\gamma$ ($=s_R$, $R=T_\gamma$), where

$$s^\beta_\gamma = -\int \frac{e^{-\gamma \Gamma}}{Z_\gamma} [(1-\beta)(-\tau_0 \omega_0) e^{\omega_0 \Gamma} + \beta(-\tau_1 \omega_1) e^{\omega_1 \Gamma}] \ln[(1-\beta)(-\tau_0 \omega_0) e^{\omega_0 \Gamma} + \beta(-\tau_1 \omega_1) e^{\omega_1 \Gamma}] \frac{d\Gamma}{\tau}. \quad (A.2)$$

Taking into account (20), we obtain that $\langle s(R) \rangle = -s^\beta_\gamma$, and expressions (1), (2), (5), (A.2) leads to (23).

Note that entropy $s_R$ characterizes the uncertainty of the parameters $R$, their statistical spread. Relationship (A.1) in [36] is interpreted as follows: the total uncertainty in the system is equal to the sum of the uncertainty of the parameters $R$ and the average uncertainty of the dynamic variables remaining after fixing the parameters $R$. Entropy $s_R$ is a negligible part of the total physical entropy due to the fact that the number of parameters $R$ is negligible small compared to the number of molecular dynamic variables $q_i$, $p_i$. The main entropy in the case of stationary nonequilibrium states is continuously produced in an open system due to the nonequilibrium nature of the processes ongoing in it and is transferred to the thermostat (the environment). In parallel, energy transfer occurs from external energy sources to the thermostat [33].